\documentclass[twocolumn,showpacs,amsmath,amssymb]{revtex4}

\usepackage{graphicx}
\usepackage{dcolumn}
\usepackage{bm}

\begin{document}


\title{Majority-vote model on random graphs}

\author{Luiz F. C. Pereira}
 \email{luizfc@df.ufpe.br}
\author{F. G. Brady Moreira}%
 \email{brady@df.ufpe.br}
\affiliation{Departamento de F\'{\i}sica, \\
Universidade Federal de Pernambuco, 50670-901, Recife-PE, Brazil}

\date{\today}

\begin{abstract}
The majority-vote model with noise on Erd\"os-R\'enyi's random graphs has been studied. 
Monte Carlo simulations were performed to characterize the order-disorder phase 
transition appearing in the system. We found that the value of the 
critical noise parameter $q_{c}$ is an increasing function of the mean 
connectivity $z$ of the random graph. The critical exponents $\beta / \nu$, 
$\gamma / \nu$ and $1 / \nu$ were calculated for several values of $z$, and our analysis yielded critical exponents satisfying the hyperscaling relation  with effective dimensionality equal to unity.
\end{abstract}

\maketitle

\section{\label{sec1}INTRODUCTION}

Since Erd\"os and R\'enyi's work more than forty years ago \cite{Erdos,Bollobas}, intense theoretical research on random graphs has been taking place \cite{Albert,Mendes}. 
In particular, models of networks with more complex connectivities than the traditional uncorrelated random graphs have been introduced \cite{WS,BA} and used to describe many systems in nature and society \cite{Manna,Vicsek,Longo}.

A random graph is a set of $N$ vertices (sites) connected by $B$ links 
(bonds). The probability $p$ that a given pair of sites is connected by a 
bond is $p=2B/N(N-1)$. The connectivity of a site is defined as the total 
number of bonds connected to it, that is $k_{i}=\sum_{j}c_{ij}$, where 
$c_{ij}=1$ if there is a link between the sites $i$ and $j$ and $c_{ij}=0$ 
otherwise. Random graphs are completely characterized by the mean number of 
bonds per site, or the average connectivity $z=p(N-1)$. In the limit 
$N \rightarrow \infty$ the distribution of connectivities is given by the 
Poisson distribution. 

For values of $z \leq 1$ the random graph does not have a percolating cluster
\cite{Bollobas, Dall}. There are a few disconnected clusters and there is no 
long-range order on such systems. For $1 < z \leq 4$ there is a percolating
cluster, but there are a few small islands disconnected from the giant cluster. These small islands do not contribute to the system dynamics and so they are excluded from our simulations. For values of $z>4$ almost all the sites belong to the giant cluster, so no sites need to be excluded from the dynamics.

Our goal in this work is to identify the critical character of the majority-vote model with noise   on random graphs. 
Previous works on the majority-vote model consider the spins either on regular d-dimensional lattices \cite{oliveira92,santos} or on small-world networks \cite{paulo,naza} interpolating between regular lattices and random graphs.  
We use Monte Carlo (MC) simulations and standard finite-size scaling techniques 
to determine the critical noise parameter $q_{c}$, as well as the exponents 
$\beta / \nu$, $\gamma / \nu$ and $1 / \nu$ for several values of the mean 
connectivity $z$ of the graph. We also use mean-field approximation to obtain the 
phase diagram in the $q_{c} - z$ space and make a comparison with the 
corresponding phase diagram that follows from our simulations.

This paper is organized in the following way: In section \ref{sec2} we describe 
the isotropic majority-vote model with noise and introduce 
the relevant quantities used in our simulations. In section \ref{sec3} we present our results 
along with a discussion. And finally, in the last section we present our 
conclusions.

\section{\label{sec2} The model and computational method}

The isotropic majority-vote model with noise is defined by a set of 
spin variables \{$\sigma_{i}$\}, where each spin is associated to one vertex of the random graph and can have the values $\pm 1$. The system dynamics 
is as follows: 
For each spin we determine the sign of the majority of its neighboring spins, that is, all the spins that are linked to it. With probability 
$q$ the spin takes the opposite sign of the majority of its neighbors, 
and it takes the same sign with probability $(1-q)$. The probability $q$ 
is known as the noise parameter.

The probability of a single-spin-flip is given by
\begin{equation}
w (\sigma _{i})=\frac{1}{2}\left[1-(1-2q)\sigma_{i} S(\sum_{\delta 
=1}^{k_i}\sigma _{i+\delta})\right],
\end{equation}
where $S(x)=\mbox{sgn}(x)$ if $x \neq 0$ and $S(0)=0$, and the summation is 
over all the $k_i$ spins connected to the spin at site $i$. 

To study the critical behavior of the model we consider the magnetization 
$M_{N}$, the susceptibility $\chi_{N}$, and the Binder's fourth-order 
cumulant $U_{N}$. These quantities are defined by
\begin{equation}
M_{N}(q) = \left\langle \left\langle m \right\rangle _{T} \right\rangle _{C} =
\left\langle \left\langle \frac{1}{N}\left| \sum_{1}^{N}\sigma_{i} \right| 
\right\rangle _{T} \right\rangle _{C},
\end{equation}
\begin{equation}
\chi_{N}(q) = N\left[ \langle ~ \langle m^{2}\rangle _{T} \rangle _{C} - \langle ~ 
\langle m \rangle _{T} \rangle _{C}
^{2}\right],
\end{equation}
\begin{equation}
U_{N}(q) = 1 - \frac{\langle ~ \langle m^4 \rangle _{T} \rangle _{C} }{3\langle ~ 
\langle m^2 \rangle _{T} \rangle _{C}^2},
\end{equation} 
where N is the number of vertices of the random graph with fixed $z$,  $\langle ... \rangle_{T}$ denotes   
time averages taken in the stationary regime, and $\langle ... \rangle_{C}$ 
stands for configurational averages.

These quantities are functions of the noise parameter $q$ and, in the critical region, satisfy the following finite-size scaling relations \cite{oliveira92}
\begin{equation} \label{mfss}
M_{N}(q) = N^{-\beta / \nu}\tilde{M}(N^{1 / \nu} \varepsilon)
\end{equation}
\begin{equation} \label{xfss} 
\chi_{N}(q) = N^{\gamma / \nu}\tilde{\chi}(N^{1 / \nu} \varepsilon)
\end{equation}
\begin{equation} \label{ufss} 
U_{N}(q) = \tilde{U}(N^{1 / \nu} \varepsilon)
\end{equation}
where $\varepsilon = q - q_{c}$.
From the size dependence of $M_{N}$ and $\chi_{N}$ we obtained the exponents 
$\beta / \nu$ and $\gamma / \nu$, respectively.
The maximum value of the susceptibility also scales as $N^{\gamma / \nu}$. 
Moreover, the value of $q$ for which $\chi_{N}$ has a maximum, $q_{c}(N)$, is expected to scale with the system size as
\begin{equation} \label{qcfss} 
q_{c}(N) = q_{c} + bN^{-1 / \nu},
\end{equation}
where the constant $b$ is close to unity. 
The above relation is used to determine the exponent $1 / \nu$ and also to provide a check for the values of $q_{c}$ obtained from the analysis of the Binder's cumulant (Eq. \ref{ufss}).
Finally, we have checked whether the calculated exponents satisfy the hyperscaling hypothesis
\begin{equation} \label{hsc}
2\beta / \nu + \gamma / \nu = D_{eff}
\end{equation}
in order to get the effective dimensionality, $D_{eff}$, 
for several values of $z$.

We have performed Monte Carlo simulations on random graphs with various values of 
mean connectivity $z$. For a given $z$, we used systems of size $N=1000, 1750, 2500, 5000$ and $10000$. We waited $8000$ Monte Carlo steps (MCS) to make the system reach the steady state, and the time averages were estimated from the next $4000$ MCS. In our simulations, one MCS is accomplished after all the $N$ spins are updated. The simulations were performed using the standard C random number generator. For all sets of parameters, we have generated ten  distinct random networks, and we have simulated ten independent runs for each distinct graph. 

\section{\label{sec3} Results and discussion}

\begin{figure}
\includegraphics[width=6cm,angle=-90]{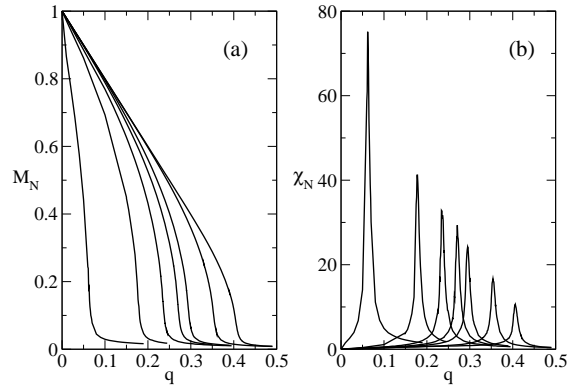}
\caption{\label{mag-susc} Magnetization and susceptibility as a function of the 
noise parameter $q$, for $N=10000$ sites. From left to right we have 
$z=2$, $4$, $6$, $8$, $10$, $20$ and $50$}
\end{figure}

\begin{figure}
\includegraphics[width=6cm,angle=-90]{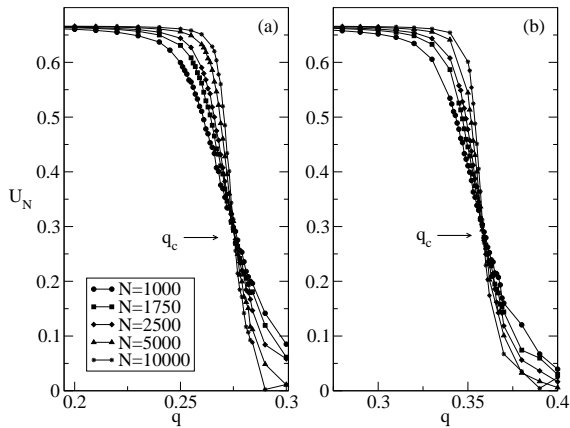}
\caption{\label{cumul} Binder's fourth-order cumulant as a function of $q$ and five values of system size $N$. In part (a) we have $z=8$ and in part (b) $z=20$.}
\end{figure}

\begin{figure}
\includegraphics[width=6cm]{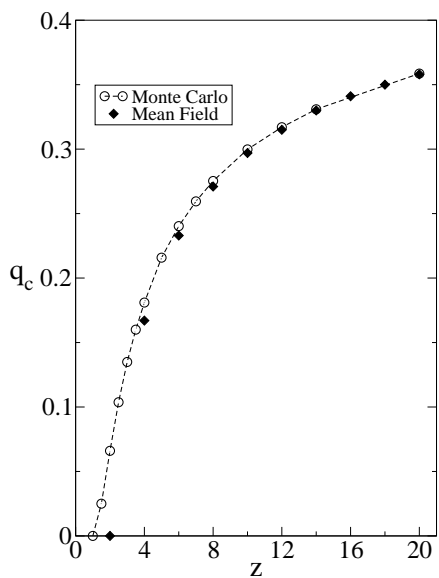}
\caption{\label{phased} The phase diagram, showing the dependence of the critical noise paramenter $q_{c}$ on the average connectivity $z$, obtained from MC simulations and from MF approximation.}
\end{figure}

In Fig. \ref{mag-susc} we show the dependence of the magnetization $M_{N}$ and the susceptibility $\chi_{N}$ on the noise parameter, obtained from simulations on random graphs with $N=10000$ nodes and several values of the average connectivity $z$.  
In part (a) each curve for $M_{N}$, for a given value of $N$ and $z$, clearly indicates 
that there exists a phase transition from an ordered state to a disordered 
state. We also notice that the transition occurs at a value of the critical 
noise parameter $q_{c}$, which is an increasing function of the mean 
connectivity $z$ of the random graph. 
In part (b) we show the corresponding behavior of susceptibilities $\chi_{N}$. The value of $q$ where $\chi_{N}$ has a maximum is here identified as $q_{c}(N)$.  
In Fig. \ref{cumul} we plot the Binder's fourth-order cumulant $U_{N}$ 
for different values of $N$ and two distinct values of $z$. 
The critical noise paramenter $q_{c}$, for a given value of $z$, 
is estimated as the point where the curves for different system sizes $N$ 
intercept each other. In this way we have obtained the phase diagram shown in Fig. \ref{phased}. 

The phase diagram of the majority-vote model on random graphs shows that for a given graph (fixed $z$) the system becomes ordered for $q<q_c$, whereas it has 
zero magnetization for $q \ge q_c$. We notice that the increase of $q_{c}$ is more pronounced for small values of $z$. The error bars in $q_{c}$ (see Table \ref{tab1}) are smaller than the symbols. 
In the figure, it is also shown the values of $q_{c}$ obtained from mean-field (MF) approximation. For small connectivities $z$ the MF estimate of the critical noise parameter is very inaccurate. In particular MF theory gives $q_c=0$ for $z \leq 2$,  whereas our MC phase diagram exhibits an ordered state for all values of the mean connectivity greater than one. This is in agreement with the limiting value of $z=1$ for the existence of a percolating cluster and, therefore, the onset of long-range order in the system. However, as $z$ increases the two estimates get closer. 
This is expected because MF approximation becomes more precise as the number of interacting nodes is increased. 
We have also performed simulations for random graphs with higher values of $z$, 
such as $z=50$, $z=100$, and $z=1000$. The corresponding values of the critical noise (not shown in the figure) are smaller than $0.50$, which is the limiting prediction value as provided by 
mean-field theory when $z \rightarrow \infty$. 

\begin{figure}
\includegraphics[width=6cm,angle=-90]{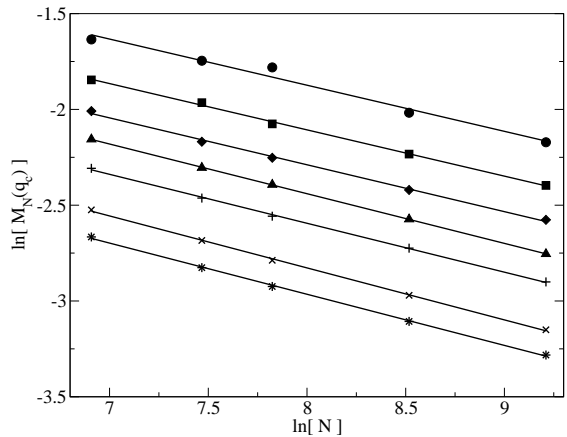}
\caption{\label{Beta}  $\ln (M_N(q_c))$ versus $\ln N$. 
From top to bottom, $z=2,4,6,10,20,50,100$.}
\end{figure}

In Fig. \ref{Beta} we plot the dependence of the magnetization at $q=q_c$ with the system size. 
The straight lines, obtained from simulations with different values of the mean connectivity $z$, confirm the scaling of the magnetization according to Eq. (\ref{mfss}). The slopes of curves correspond to the exponent ratio $\beta / \nu$. Our results show that the increase of $\beta / \nu$ with $z$ is quite small.  

\begin{figure}
\includegraphics[width=6cm,angle=-90]{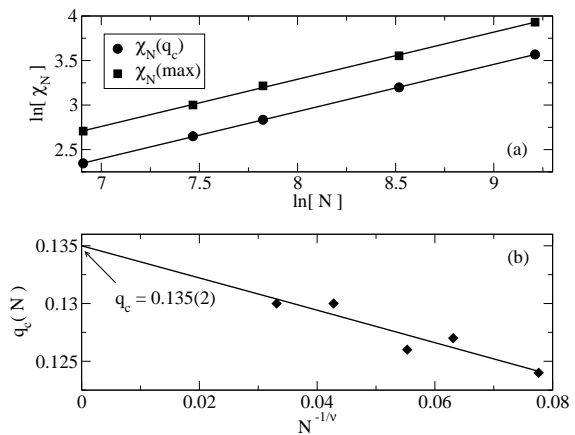}
\caption{\label{Susc-nu} (a) Plot of $\ln (\chi_{N}(q_{c}))$ and $\ln (\chi_{N}(max))$ versus $\ln N$. (b) The dependence of the noise parameter $q_c(N)$ on system size. The extrapolation gives an independent estimation for $q_c$.
The data are for the case of mean connectivity $z=3$.}
\end{figure}

In Fig. \ref{Susc-nu}(a) we display the scalings for the susceptibility at $q=q_c$, 
$\chi_N(q_c)$, and for its maximum amplitude, $\chi_N(max)$. The exponent ratio  $\gamma / \nu$ are obtained from the slopes of the straigth lines. For almost all the values of $z$, the exponents $\gamma / \nu$ of the two estimates agree within error bars (Table \ref{tab1}). An increased $z$ means a slight tendency to decrease the exponent ratio $\gamma / \nu$.  

In a similar way, for fixed $z$ the critical exponent $1/ \nu$ was obtained from a plot of $\ln {q_{c}(N)- q_c}$ versus $\ln N$ (see Eq. (\ref{qcfss})). 
We used the corresponding values of $q_c(N)$ that follow from the maximum of the susceptibility and the limiting value of $q_c$ which has been obtained from Binder's cumulant. The slope of the resulting straight line equals the exponent $1/ \nu$. The results quoted in Table \ref{tab1} indicate that $1/ \nu$ is not a monotonic function of the mean connectivity $z$. 

Fig. \ref{Susc-nu}(b) illustrates the scaling relation for $q_c(N)$ given in Eq. (\ref{qcfss}).  
The constant $b$ equals the slope of the straigth line, whereas the 
extrapolation of the fitting provides an alternative way of determining 
the critical parameter $q_{c}$. We have obtained a quite satisfactory agreement between the values of $q_{c}$ determined in this way and the corresponding ones that follow from the analysis of Binder's cumulant.

\begin{figure}
\includegraphics[width=6cm,angle=-90]{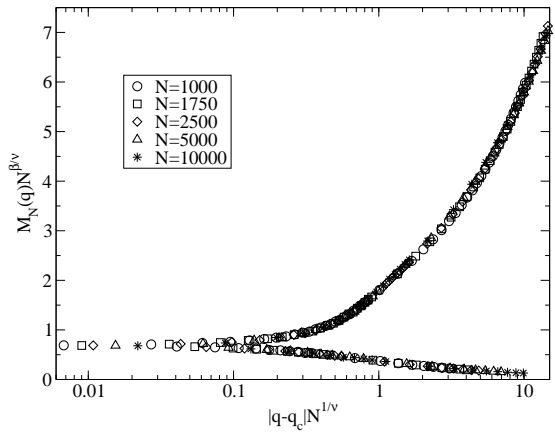}
\caption{\label{datacol} Data-collapsing for five different values of $N$, with $z=10$.}
\end{figure}

In Fig. \ref{datacol} we show the data-collapse plot for 
$\tilde{M}(u)=M_{N}(q) N^{\beta / \nu}$, which is a universal function of 
the combined variable $u=N^{1 / \nu} (q-q_c)$. We have also obtained quite good 
data-collapse for $\tilde{\chi}(u)=\chi_{N}(q) N^{-\gamma / \nu}$. 
The collapsing of curves for five different system sizes corroborates the quoted values for $q_c$, $\beta / \nu$, $\gamma / \nu$ and $1 / \nu$. 

Table \ref{tab1} resumes the values (along with errors) of 
$q_{c}$, the three critical exponents ($\gamma / \nu$ was obtained by using two different scalings), and the effective dimensionality of the system.
It is worth to mention that, for all $z$, the value $D_{eff}=1$, which has been obtained from the hyperscaling hypothesis (Eq. (\ref{hsc})), is satisfied when we use both estimate procedures for the exponent ratio $\gamma / \nu$. 

As far as we know, there is no previous works studying the majority-vote model on Erd\"os-R\'enyi's graphs, to allow a direct comparison of the present results. 
Yet, for completeness, it would be of interest to mention earlier simulations of the majority-vote model on other kinds of networks. Campos et al \cite{paulo} investigated  the phase diagram and critical behavior of the majority-vote model on small-world networks \cite{WS} by rewiring the two dimensional square lattice. Using a similar procedure to ours they found critical exponents 
depending on the fraction of long-range interactions and satisfying the hiperscaling relation with $D_{eff}=2$ (the dimensionality of the regular lattice). On the other hand, the model which has been defined on a regular lattice has critical 
exponents that fall into the same class of universality as the corresponding equilibrium Ising model \cite{oliveira92,santos}. 
The results of the present simulations show that the majority-vote models defined on a regular lattice, on small-world networks, and on Erd\"os-R\'enyi's random graphs belong to different universality classes. 

Finally, we remark that our MC results are quite different from the mean-field estimates $\beta / \nu=1$, $\gamma / \nu=2$ and $1 / \nu=2$, which result in $D=4$ for the upper critical dimensionality. This is a reasonable result since for all networks simulated we are far away from the mean-field picture where every spin interacts with all the remaining $N-1$ spins. 

\begin{table}
\caption{\label{tab1} The critical noise $q_c$, the critical exponents, and the effective dimensionality $D_{eff}$, for random networks with mean connectivity $z$.}
\begin{ruledtabular}
\begin{tabular}{ccccccc}
z & $q_{c}$ & $\beta / \nu$ & $\gamma / \nu$ \footnotemark[1] & $\gamma / \nu$ \footnotemark[2] & 
$1 / \nu$ & $D_{eff}$ \footnotemark[3] \\
\hline
2 & 0.066(1) & 0.24(2) & 0.51(3) & 0.52(2) & 0.48(5) & 0.99(7) \\
3 & 0.1349(6) & 0.233(2) & 0.529(2) & 0.529(7) & 0.37(9) & 1.00(1) \\
4 & 0.181(1) & 0.242(6) & 0.54(1) & 0.515(6) & 0.59(7) & 1.02(2) \\
6 & 0.2403(5) & 0.245(7) & 0.514(4) & 0.507(4) & 0.44(6) & 1.00(2) \\
8 & 0.2753(3) & 0.242(5) & 0.510(9) & 0.510(3) & 0.56(3) & 0.99(2) \\
10 & 0.2998(4) & 0.259(1) & 0.483(5) & 0.502(5) & 0.51(3) & 1.00(1) \\
20 & 0.3586(2) & 0.255(4) & 0.501(6) & 0.503(2) & 0.49(4) & 1.01(1) \\
50 & 0.4110(2) & 0.271(4) & 0.465(6) & 0.485(4) & 0.39(5) & 1.01(1) \\
100 & 0.4368(3) & 0.267(4) & 0.467(7) & 0.479(4) & 0.47(3) & 1.00(2) \\
\end{tabular}
\end{ruledtabular}
\footnotetext[1]{Obtained using $\chi_N(q_c)$. See Eq. (\ref{xfss}).}
\footnotetext[2]{Obtained using $\chi_N(max)$.}
\footnotetext[3]{Estimated using $\gamma / \nu$ from $\chi_N(q_c)$.}
\end{table}

\section{Conclusion} 

We have obtained the phase diagram and critical exponents of the majority-vote model 
with noise on random graphs. The second-order phase transition which occurs in the model with mean connectivity $z>1$ has exponents that show a slight variation along the critical line. Nevertheless, our Monte Carlo simulations have demonstrated that the effective dimensionality $D_{eff}$ equals unity, for all values of $z$. This interesting result may suggest that other spin models defined on random graphs have exponents which satisfy the hyperscaling relation with $D_{eff}=1$. 

\begin{acknowledgments}
Luiz F. C. Pereira is supported by CAPES. We acknowledge partial 
support from CNPq, FINEP and FACEPE. 
\end{acknowledgments}


\end{document}